# Optimising data for modelling neuronal responses

Peter Zeidman[1*+], Samira M Kazan[1+], Nick Todd[2], Nikolaus Weiskopf[1,3], Karl J. Friston[1], and Martina F. Callaghan[1]

[1]*Wellcome Centre for Human Neuroimaging, UCL Institute of Neurology, University College London, London, United Kingdom*

[2]*Department of Radiology, Brigham and Women's Hospital, Harvard Medical School, Boston, Massachusetts, United States*

[3]*Department of Neurophysics, Max Planck Institute for Human Cognition and Brain Sciences, Leipzig, Germany*

+Equal contribution
*Corresponding author

# Abstract

In this technical note, we address an unresolved challenge in neuroimaging statistics: how to determine which of several datasets is the best for inferring neuronal responses. Comparisons of this kind are important for experimenters when choosing an imaging protocol – and for developers of new acquisition methods. However, the hypothesis that one dataset is better than another cannot be tested using conventional statistics (based on likelihood ratios), as these require the data to be the same under each hypothesis. Here we present Bayesian data comparison, a principled framework for evaluating the quality of functional imaging data, in terms of the precision with which neuronal connectivity parameters can be estimated and competing models can be disambiguated. For each of several candidate datasets, neuronal responses are inferred using Dynamic Casual Modelling (DCM) – a commonly used Bayesian procedure for modelling neuroimaging data. Next, the parameters from subject-specific models are summarised at the group level using a Bayesian General Linear Model (GLM). A series of measures, which we introduce here, are then used to evaluate each dataset in terms of the precision of (group-level) parameter estimates and the ability of the data to distinguish similar models. To exemplify the approach, we compared four datasets that were acquired in a study evaluating multiband fMRI acquisition schemes. To enable people to reproduce these analyses using their own data and experimental paradigms, we provide general-purpose Matlab code via the SPM software.

# Introduction

Hypothesis testing involves comparing the evidence for different models or hypotheses, given some measured data. The key quantity of interest is the likelihood ratio - the probability of observing the data under one model relative to another – written $p(y|m_1)/p(y|m_2)$ for models $m_1$ and $m_2$ and dataset $y$. Likelihood ratios are ubiquitous in statistics, forming the basis of the F-test and the Bayes factor in classical and Bayesian statistics respectively. They are the most powerful test for any given level of significance by the Neyman-Pearson lemma (Neyman, 1933). However, the likelihood ratio test assumes that there is only one dataset $y$ – and so cannot be used to compare different datasets. An unresolved problem, especially pertinent to neuroimaging, is how to test the hypothesis that one dataset is better than another for making inferences.

Neuronal activity and circuitry cannot generally be observed directly, but rather are inferred from measured timeseries. In the case of fMRI, the data are mediated by neuro-vascular coupling, the BOLD response and noise. To estimate the underlying neuronal responses, models are specified which formalise the experimenter's understanding of how the data were generated. Hypotheses are then tested by making inferences about model parameters, or by comparing the evidence under different models. For example, an F-test can be performed on the General Linear Model (GLM) using classical statistics, or Bayesian Model Comparison can be used to select between probabilistic models. From the experimenter's perspective, the best dataset provides the most precise estimates of neuronal responses (enabling efficient inference about parameters) and provides the greatest discrimination among competing models (enabling efficient inference about models).

Here, we introduce Bayesian data comparison – a series of information measures for evaluating a dataset's ability to support inferences about both parameters and models. While they are generic and can be used with any sort of probabilistic models, we illustrate their application using Dynamic Causal Modelling (DCM) for fMRI (Friston et al., 2003) because it offers several advantages. Compared to the GLM, DCM provides a richer characterisation of neuroimaging data, through the use of biophysical models, based on differential equations that separate neuronal and haemodynamic parameters. This means one can evaluate which dataset is best for estimating neuronal parameters specifically. These models also include connections among regions; making DCM the usual approach for inferring effective (directed) connectivity from fMRI data.

By using the same methodology to select among datasets as experimenters use to select between connectivity models, feature selection and hypothesis testing can be brought into alignment for connectivity studies. Moreover, a particularly useful feature of DCM for comparing datasets is that it employs Bayesian statistics. The posterior probability over neuronal parameters forms a multivariate normal distribution, providing expected values for each neuronal parameter, as well as their covariance. The precision (inverse covariance) quantifies the confidence we place in the parameter estimates, given the data. We will use this precision to assess the quality of parameter estimates in different datasets (under the assumption that the experimental effect does or does not exist, based on previous experiments). DCM also enables experimenters to distinguish among models, in terms of which model maximises the log model evidence $\ln p(y|m)$. We cannot use this to compare different datasets, but we can ask which of several datasets provides the most efficient discrimination among models. For these reasons, we used DCM as the basis for comparing datasets, both to provide estimates of neuronal responses and to distinguish among neural models.

This paper presents a methodology and associated software for evaluating which of several imaging acquisition protocols or data features affords the most sensitive inferences about neural architectures. We illustrate the framework by assessing the quality of fMRI time series acquired from ten participants, who were each scanned 4 times with a different multiband acceleration factor. Multiband is an approach for rapid acquisition of fMRI data, in which multiple slices are acquired simultaneously and subsequently unfolded using coil sensitivity information (Larkman et al., 2001; Xu et al., 2013). The rapid sampling enables sources of physiological noise to be separated from sources of experimental variance more efficiently. However, a penalty for this increased temporal resolution is a reduction of the signal-to-noise ratio (SNR) of each image. A detailed analysis of these data is published separately (Todd et al., 2017). We do not seek to draw any novel conclusions about multiband acceleration from this specific dataset, but rather to use it to illustrate a generic approach for comparing datasets.

The methodology we introduce here offers several novel contributions. First, it provides a sensitive comparison of data by evaluating their ability to reduce uncertainty about neuronal parameters, and to discriminate among competing models. Second, our procedure identifies the best dataset for hypothesis testing at the group level, reflecting the objectives of most cognitive neuroscience studies. Unlike a classical GLM analysis – where only the maximum likelihood estimates from each subject are taken to the group level – the (parametric empirical) Bayesian methods used here take into account the uncertainty of the parameters (the full covariance matrix), when modelling at the group level. Additionally, this methodology provides the necessary tools to evaluate which imaging protocol is optimal for effective connectivity analyses; although we anticipate many questions about data quality will not necessarily relate to connectivity. We provide a single Matlab function for conducting all the analyses described in this paper, which is available in the SPM (http://www.fil.ion.ucl.ac.uk/spm/) software package (**spm_dcm_bdc.m**). This function can be used to evaluate any type of imaging protocol, in terms of the precision with which model parameters are estimated and the complexity of the generative models that can be disambiguated.

## Methods

We begin by briefly reprising the theory behind DCM and introducing the series of outcome measures used to evaluate data quality. We then move on to illustrating the analysis pipeline in the context of an exemplar fMRI dataset.

### Dynamic Causal Modelling

DCM is a framework for evaluating generative models of time series data. At the most generic level, neuronal activity in region $i$ of the brain at time $t$ may be modelled by a lumped or neural mass quantity $z_t^{(i)}$. Generally, the experimenter is interested in the neuronal activity of a set of interconnected brain regions, the activity of which can be written as a vector $z(t)$. The evolution of neural activity over time can then be written as:

$$\dot{z} = f(z, u, \theta^{(n)}) \qquad (1$$

Where $\dot{z}$ is the derivative of the neural activity with respect to time, $u$ are the time series of experimental or exogenous inputs and $\theta^{(n)}$ are the parameters controlling connectivity within and between regions. Neural activity cannot generally be directly observed. Therefore, the neural model is combined with an observation model $g$, with parameters $\theta^{(h)}$, which specifies how neural activity is transformed into timeseries:

$$y = g(z, \theta^{(h)}) + \epsilon^{(1)} \qquad (2$$

Where $\epsilon^{(1)}$ is zero-mean (I.I.D.) additive Gaussian noise, with log-precision specified by hyperparameters $\lambda = \lambda_1 \ldots \lambda_r$ for each region $r$. The I.I.D. assumption is licensed by automatic pre-whitening of the timeseries in SPM, prior to the DCM analysis. In practice, it is necessary to augment the vector of neuronal activity with haemodynamic parameters that enter the observation model above. The specific approximations of functions $f$ and $g$ depend on the imaging modality being used, and the procedures described in this paper are not contingent on any specific models. However, to briefly reprise the basic model for fMRI – which we use here for illustrative purposes – $f$ is a Taylor approximation to any nonlinear neuronal dynamics:

$$\dot{z} = \left(A + \sum u_j B^{(j)}\right) z + Cu \quad (3$$

There are three sets of parameters $\theta^{(n)} = (A, B, C)$ and $j$ experimental inputs. Matrix $A$ represents the strength of connections within (i.e., intrinsic) and between (i.e. extrinsic) regions - their effective connectivity. Matrix B represents the effect of time-varying experimental inputs on each connection (these are referred to as modulatory or condition-specific effects). Matrix $C$ specifies the influence of each experimental input on each region, which effectively drives the dynamics of the system. The vector $u_j$ is a timeseries encoding the timing of experimental condition $j$.

The haemodynamics (the observation model $g$ above) are modelled with an extended Balloon model (Buxton et al., 2004; Stephan et al., 2007), which comprises a series of differential equations describing the process of neurovascular coupling by which activity ultimately manifests as a BOLD signal change. The majority of the parameters of this haemodynamic model are based on previous empirical measurements; however, three parameters are estimated on a region-specific basis: the transit time $\tau$, the rate of signal decay $\kappa$, and the ratio of intra- to extra-vascular signal $\epsilon^{(h)}$.

The observation parameters $\theta^{(h)}$ are concatenated with the neural parameters $\theta^{(n)}$ and the hyperparameters $\lambda$ and these are estimated using a standard variational Bayes scheme called variational Laplace (Friston et al., 2003; Friston, 2011). This provides a posterior probability density for the parameters, as well as an approximation of the log model evidence (i.e., the negative variational free energy), which scores the quality of the model in terms of its accuracy minus its complexity.

### Group analyses with PEB

Having fitted a model of neuronal responses to each subject individually (a first level analysis), the parameters can be summarised at the group level (a second level analysis). We used a Bayesian GLM, implemented using the Parametric Empirical Bayes (PEB) framework for DCM (Friston et al., 2016). With $N$ subjects and $M$ connectivity parameters for each subject's DCM, the group-level GLM has the form:

$$\theta = X\beta + \epsilon^{(2)} \quad (4$$

The dimensions of this GLM are illustrated in Figure 1. Vector $\theta \in \mathbb{R}^{NM \times 1}$ are the neuronal parameters from all the subjects' DCMs, consisting of all parameters from subject 1, then all parameters from subject 2, etc. The design matrix $X \in \mathbb{R}^{NM \times M}$ was specified as:

$$X = \mathbf{1}_N \otimes I_M \quad (5$$

Where $\mathbf{1}_N$ is a column vector of 1s of dimension $N$ and $I_M$ is the identity matrix of dimension $M$. The Kronecker product $\otimes$ replicates the identity matrix vertically for each subject. The use of a Kronecker product at the between subject level reflects the fact that between subject effects can be expressed at each and every connection. In this instance, we are just interested in the group mean and therefore there is only one between-subject effect. The resulting matrix $X$ has one column (also called a covariate or regressor) for each connectivity parameter (Figure 1). The regressors are scaled by parameters $\beta \in \mathbb{R}^{M \times 1}$, which are estimated from the data and represent the group average strength of each connection. Finally, the errors $\epsilon^{(2)} \in \mathbb{R}^{NM \times 1}$ are modelled as zero-mean additive noise:

$$\epsilon^{(2)} = N(0, \Pi^{-1}) \quad (6$$

Where precision matrix $\Pi \in \mathbb{R}^{NM \times NM}$ is estimated from the data. This captures the between-subject variability in the connection strengths, parameterised using a single parameter $\gamma$:

$$\Pi = I_N \otimes (Q_0 + e^{-\gamma} Q_1) \quad (7$$

This is a multi-component covariance model. $Q_0 \in \mathbb{R}^{M \times M}$ is the lower bound on precision and ensures it is a positive number. Matrix $Q_1 \in \mathbb{R}^{M \times M}$ is the prior precision. When the parameter $\gamma$ is zero, the precision is equal to $Q_0 + Q_1$. More negative values of $\gamma$ equate to higher precision than the prior and *vice versa*. The Kronecker product $\otimes$ replicates the precision matrix for each subject, giving rise to the matrix $\Pi$ of dimension $NM \times NM$ where the leading diagonal is the precision of each DCM parameter $\theta$.

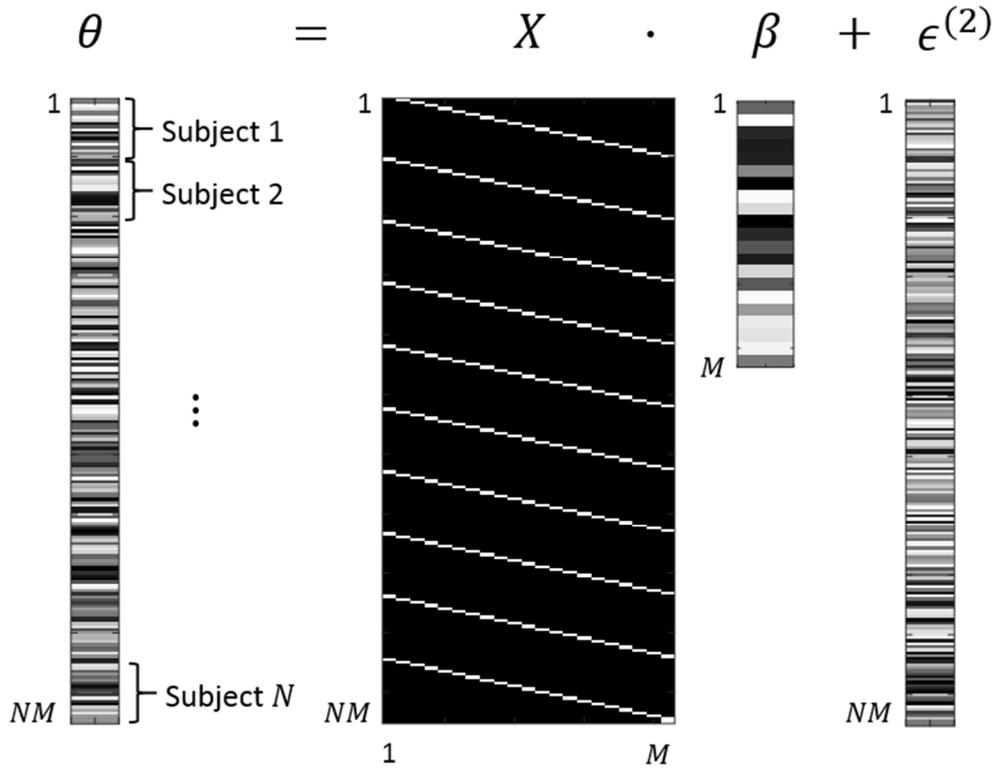

**Figure 1 Form of the General Linear Model (GLM).** The $M$ parameters from all $N$ subjects' DCMs are arranged in a vector $\theta$. This is modelled using a design matrix $X$ which encodes which DCM parameter is associated with which element of $\theta$. After estimation of the GLM, parameters $\beta$ are the group average of each DCM connection. Between-subjects variability $\epsilon^{(2)}$ is specified according to Equation 6. In this figure, white=1 and black=0. Shading in parameters $\theta$, $\beta$, $\epsilon$ is for illustrative purposes only.

The group-level parameters representing the average connection strengths across subjects $\beta$ and the parameter controlling between-subject variability $\gamma$ are specified as normal distributions, and the PEB scheme provides estimates of their expected values and covariance (uncertainty):

$$\beta = N(\mu_\beta, \Sigma_\beta)$$
$$\gamma = N(\mu_\gamma, \Sigma_\gamma)$$

(8

Where $\mu_\beta \in \mathbb{R}^{MN \times 1}, \Sigma_\beta \in \mathbb{R}^{MN \times MN}, \mu_\gamma \in \mathbb{R}^1, \Sigma_\gamma \in \mathbb{R}^1$.

To summarise, within and between-subject parameters are estimated using a hierarchical scheme, referred to as Parametric Empirical Bayes (PEB). Model estimation provides the approximate log model evidence (free energy) of the group-level Bayesian GLM – a statistic that enables different models to be compared. The free energy can be decomposed into the accuracy minus the complexity of the model, where the complexity is defined by the KL-divergence between the posterior and prior parameters. We take advantage of the free energy below to compare models of group-level data. For full details on the priors, model specification and estimation procedure in the PEB scheme, see Friston et al. (2016).

Readers familiar with Random Effects Bayesian Model Selection (RFX BMS) (Stephan et al., 2009) will note the distinction with the PEB approach used here. Whereas RFX BMS considers random effects over models, the PEB approach considers random effects at the group level to be expressed at the level of parameters; namely, parametric random effects. This means that uncertainty about parameters at the subject level is conveyed to the group level; licensing the measures described in the next section.

## Outcome measures

### a) Parameter certainty

To measure the information gain (or reduction in uncertainty) about parameters, we take advantage of the Laplace approximation used in the DCM framework, which means that the posterior and prior densities over parameters are Gaussian. In this case, the confidence of the parameter estimates can be quantified using the negative entropy of the posterior multivariate density over interesting parameters:

$$S_\theta = -0.5 \ln |2\pi e \, \Sigma_\beta| \qquad (9$$

Equation 9 uses the definition of the negative entropy for the multivariate normal distribution, applied to the neuronal parameter covariance matrix $\Sigma_\beta$. This has units of nats (i.e., natural units) and provides a summary of the precision associated with group level parameters – such as group means – having properly accounted for measurement noise and random effects at the between-subject level.

Datasets can be compared by treating the entropies as log Bayes factors (detailed in *Appendix 1: Bayesian data comparison*). In brief, this follows because the log Bayes factor can always be decomposed into two terms – a difference in accuracy minus a difference in complexity. The complexity is the KL-divergence between the posteriors $\Sigma_\beta$ and the priors $\Sigma_0$, and it scores the reduction in uncertainty afforded by the data, in units of nats. Under flat or uninformative priors, the KL-divergence reduces to the negative entropy in Equation 9. A difference in entropy of 3 nats corresponds to $e^3 \approx 20$ times the information gain, and corresponds to 'strong evidence' (Kass and Raftery, 1995) that one dataset is more informative than another. These same principles apply for the following three measures below.

### b) Random effects certainty

In addition to the connectivity parameters, the PEB model also provides estimates of the between-subject variability. In statistics, this is known as random effects. The precision term $\Pi$ (Equation 6) reflects any unmodelled sources of variation between subjects, and our uncertainty about $\Pi$ is represented by the variance parameter $\Sigma_\gamma$ which is estimated from the data (Equation 8). We can quantify this uncertainty about random effects using the negative entropy:

$$S_\epsilon = -0.5 \ln|2\pi e\, \Sigma_\gamma| \qquad (10$$

Although we have used a single precision component to model between-subject variability ($Q_1$ in Equation 7), this can be generalised to multiple components, for example, to enable different levels of between-subject variability in different parts of the model. In this general case, $\Sigma_\gamma$ will be a matrix rather than a scalar.

### c) Information gain (parameters)

The data quality afforded by a particular acquisition scheme can be scored in terms of the *relative entropy* or *Kullback-Leibler divergence* (Kullback and Leibler, 1951) between posterior and prior distributions over parameters. This measure of salience is also known as Bayesian surprise, epistemic value or information gain and can be interpreted as the quantitative reduction of uncertainty after observing the data. In other words, it reflects the complexity of the model (the number of independent parameters) that can be supported by the data. This takes into account both the posterior expectation and precision of the parameters relative to the priors, whereas the measure in part (a) considered only the posterior precision (relative to uninformative priors).

The KL-divergence for the multivariate normal distribution, between the posterior $N_1$ and the prior $N_0$, with mean $\mu_1$ and $\mu_0$ and covariance $\Sigma_1$ and $\Sigma_0$ respectively, is defined as:

$$D_{KL}(N_1 \parallel N_0) = \frac{1}{2}\left(\mathrm{tr}(\Sigma_0^{-1}\Sigma_1) + (\mu_0 - \mu_1)^T \Sigma_0^{-1}(\mu_0 - \mu_1) - k + \ln\frac{\det \Sigma_1}{\det \Sigma_0}\right) \qquad (11$$

Where $k = rank(\Sigma_0)$. This statistic increases when the posterior mean has moved away from the prior mean or when the precision of the parameters has increased relative to the precision of the priors. Note that this same quantity also plays an important role in the definition of the free energy approximation to log model evidence, which can be decomposed into accuracy minus complexity, the latter being the KL-divergence between posteriors and priors.

The measures described so far are based on posterior estimates of model parameters. We now turn to the equivalent measures of posterior beliefs about the models *per se*.

### d) Information gain (models)

The quality of the data from a given acquisition scheme can also be assessed in terms of their ability to reduce uncertainty about models. This involves specifying a set of equally plausible, difficult to disambiguate models that vary in their connectivity structures, and evaluating which dataset best enables these models to be distinguished.

Bayesian model comparison starts with defining a prior probability distribution over the models $P_0$. Here, we assume that all models are equally likely, therefore $P_0 = 1/p$ for each of $p$ models. This prior is combined with the model evidence, to provide a posterior distribution over the models, $P$. To quantify the extent to which the competing models have been distinguished from one another, we measure the information gain from the prior $P_0$ to the posterior $P$. This is given by the KL-divergence used above for the parameters. After describing how we specify these models, we provide an example of this KL-divergence in practice.

To construct a set of models we adopt the following procedure. We first estimate a 'full' group-level Bayesian GLM with all relevant free parameters from the subjects' DCMs. Next, we identify a set of reduced GLMs that only differ slightly in log evidence (i.e. they are difficult to discriminate). To do this we eliminate one connection or parameter (by fixing its prior variance to zero) and retain the model if the change in log evidence is greater than -3. This corresponds to a log odds ratio of approximately one in $e^3 \approx 20$; meaning that the model is retained if it is no more than 20 times less probable than the full model. We repeat this procedure by eliminating another parameter (with replacement), ultimately obtaining the final model space. This procedure can be performed rapidly by using Bayesian Model Reduction (BMR), which analytically computes the log evidence of reduced models from a full model (Friston et al., 2016).

Having identified a set of plausible but difficult to disambiguate models (GLMs) for a given dataset, we then calculate the posterior probability of each model. Under flat priors, this is simply the softmax function of the log model evidence, as approximated by the free energy (see Appendix 1). We then compute the KL-divergence between the posterior and prior model probabilities, which is defined for discrete probability distributions as:

$$D_{KL}(P \parallel P_0) = \sum_{i=1\ldots k} (P_i \ln P_i) + \ln k \qquad (12$$

The behaviour of the KL-divergence is illustrated in Figure 2, when comparing $k = 10$ simulated models. When one model has a posterior probability approaching one, and all other models have probability approaching zero, the KL-divergence is maximised and has the value $D_{KL} = \ln k = 2.30$ (Figure 2A). As the probability density is shared between more models, so the KL-divergence is reduced (Figure 2B-2C). It reaches its minimum value of zero when all models are equally likely, meaning that no information has been gained by performing the model comparison (Figure 2D).

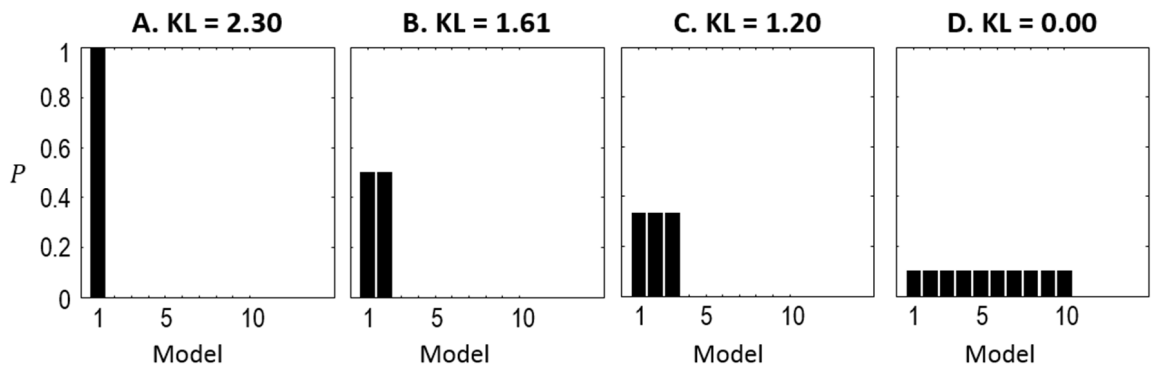

**Figure 2 Illustration of the KL-divergence in four simulated model comparisons**. The bars show the posterior probabilities of 10 models and the titles give the computed KL-divergence from the priors. **A**. Model 1 has posterior probability close to 1. The KL-divergence is at its maximum of $\ln 10 = 2.3$. **B**. The probability density is shared between models 1 and 2, reducing the KL-divergence. **C**. The probability density shared between 3 models. **D**. The KL-divergence is minimized when all models are equally likely, meaning no information has been gained relative to the prior.

## Summary of measures & analysis pipeline

A key contribution of the measures introduced in this paper is the characterisation of information gain in terms of both the parameters, and the models that entail those parameters. Together they provide a principled means by which to characterise the optimality of a given scheme for acquiring data. These are intended for use where the presence or absence of experimental effects in the data is already known – for example, based on previous studies and / or the results or an initial analysis collapsed across datasets (e.g. a mass-univariate GLM analysis).

We now suggest a pipeline for applying these measures to neuroimaging data. Step 1 provides estimates of neuronal parameters from each dataset. Steps 2 and 3 use the estimated parameters from all datasets to automatically identify a suitable model architecture (and could be skipped if the experimenter has strong priors as to what the model architecture should be). Steps 4 and 5 provide estimates of the group level parameters for each dataset and compare them using the measures described above. For convenience, steps 2-5 of this procedure can be run with a single Matlab function implemented in SPM (**spm_dcm_bdc.m**):

1) **Model each subject's data using a Bayesian model (e.g. DCM)**. The objective is to obtain posterior estimates of neuronal parameters from each dataset. These estimates take the form of a multivariate probability density for each subject and dataset.
2) **Identify the optimal group-level model structure**. This step identifies a parsimonious model architecture, which can be used to model all datasets. To achieve this, specify a Bayesian GLM and fit it to the neuronal parameters from all subjects and datasets. To avoid bias, do not inform the GLM that the data derive from multiple datasets. The estimated GLM parameters represent the average connectivity across all datasets. Prune this GLM to remove any redundant parameters (e.g. relating to the responses of specific brain regions) that do not contribute to the model evidence, using Bayesian Model Reduction (Friston et al., 2016). This gives the optimal reduced model structure at the group level, agnostic to the dataset.
3) **Re-estimate each subject's individual DCM** having switched off any parameters that were pruned in step 2. This step equips each subject with a parsimonious model to provide estimates of neuronal responses. This is known as 'empirical Bayes', as the priors for the individual subjects have been updated based on the group level data. Again, this is performed analytically using Bayesian Model Reduction.
4) **Fit a separate Bayesian GLM to the neuronal parameters of each dataset**. This summarises the estimated neuronal responses for each dataset, taking into account both the expected values and uncertainty of each subject's parameters.
5) **Apply the measures outlined above** to compare the quality or efficiency of inferences from each dataset's Bayesian GLM – in terms of parameters or models.

Collectively, the outcome measures that result from this procedure constitute an assessment of the goodness of different datasets in terms of inferences about connection parameters and models. Next, we provide an illustrative example using empirical data from an experiment comparing different fMRI multiband acceleration factors.

## Multiband example

For this example, we use fMRI data from a previously published study that evaluated the effect of multiband acceleration on fMRI data (Todd et al., 2017). First, to briefly reprise the objectives of that study. For a given effect size of interest, the statistical power of an fMRI experiment can be improved by acquiring a greater number of sample points (i.e., increasing the efficiency of the design) or by reducing measurement noise. This has the potential to enable more precise parameter estimates and provide support for more complex models of how the data were generated. Acquiring data with high temporal resolution both increases the number of samples per unit time and allows physiologically-driven fluctuations in the time series to be more fully sampled and subsequently removed or separated from the task-related BOLD signal (Todd et al., 2017). One approach to achieving rapid acquisitions is the use of the multiband or simultaneous multi-slice acquisition technique (Setsompop et al., 2012; Xu et al., 2013); in which multiple slices are acquired simultaneously and subsequently unfolded using coil sensitivity information (Setsompop et al., 2012; Cauley et al., 2014). The penalty for the increased temporal resolution is a reduction of the signal-to-noise ratio (SNR) of each image. This is caused by increased g-factor penalties, dependent on the coil sensitivity profiles, and reduced steady-state magnetisation arising from the shorter repetition time (TR) and concomitant reduction in excitation flip angle. In addition, a shorter TR can be expected to increase the degree of temporal auto-correlation in the time series. This raises the question of which MB acceleration factor offers the best trade-off between acquisition speed and image quality.

We do not seek to resolve the question of which multiband factor is optimal in general. Furthermore, there are many potential mechanisms by which multiband acquisitions could improve or limit data quality; including better sampling of physiological noise and increasing the number of samples in the data. Rather than trying to address these questions here, we instead describe an exemplar analysis to illustrate the framework for comparing datasets. In these data, physiologically-driven fluctuations – that are better sampled with higher multiband acceleration factor due to the higher Nyquist sampling frequency – were removed from the data by filtering. Subsequently, the data were down-sampled so as to have equivalent numbers of samples across multiband factors, as described in (Todd et al., 2017). The framework presented here could be used to test the datasets under many different acquisition and pre-processing procedures.

Data acquisition

Ten healthy volunteers were scanned with local ethics committee approval on a Siemens 3T Tim Trio scanner. For each volunteer, fMRI task data with 3mm isotropic resolution were acquired four times with a MB factor of either 1, 2, 4 or 8 using the gradient echo EPI sequence from the Center for Magnetic Resonance Research (R012 for VB17A, https://www.cmrr.umn.edu/multiband/). The TR was 2800ms, 1400ms, 700 ms and 350 ms for MB factor 1, 2, 4 and 8 respectively resulting in 155, 310, 620 and 1240 volumes respectively, leading to a 7 and a half minute acquisition time per run. The data were acquired with the blipped-CAIPI scheme (Setsompop et al., 2012), without in-plane acceleration, and the leak-block kernel option for image reconstruction was enabled (Cauley et al., 2014).

## fMRI task

The fMRI task consisted of passive viewing of images, with image stimuli presented in 8s blocks. Each block consisted of 4 images of naturalistic scenes or 4 images of single isolated objects, displayed successively for 2s each. There were two experimental factors: stimulus type (images of scenes or objects) and novelty (2, 3 or 4 novel images per block, with the remainder repeated). This paradigm has previously been shown to induce activation in a well-established network of brain regions that respond to perceiving, imagining or recalling scenes (Spreng et al., 2009; Zeidman et al., 2015).

## Preprocessing

All data were processed in SPM (Ashburner and Friston, 2005), version 12. This comprised the usual image realignment, co-registration to a T1-weighted anatomical image and spatial normalization to the Montreal Neurological Institute (MNI) template space; using the unified segmentation algorithm, and smoothing with a 6 x 6 x 6 mm full width at half maximum (FWHM) Gaussian kernel.

As described in Todd et al. (2016), all data were filtered using a $6^{th}$-order low pass Butterworth filter with a frequency cut-off of 0.18 Hz (corresponding to the Nyquist frequency of the MB=1 data). This removed all frequency components between the cut-off frequency and the corresponding Nyquist frequency of the particular MB factor. In order to ensure equal numbers of samples per data set – regardless of MB factor used – the time series were decimated by down-sampling all datasets to the TR of the MB1 (TR=2.8s) data.

After initial processing and filtering, all data sets were modelled with a general linear model (GLM), with a high pass filter (cut-off period=128s) and regressors for motion and physiological effects. In addition to these confounding effects, the stimulation blocks were modelled with boxcar functions convolved with the canonical haemodynamic response function. Temporal autocorrelations were accounted for with an autoregressive AR(1) model plus white noise. This was deemed sufficient given that after filtering and decimation each time series had an effective TR of 2.8s. The contrast of scenes>objects was computed and used to select brain regions for the DCM analysis.

## DCM specification

We selected seven brain regions (Figure 3A) from the SPM analysis which are part of a 'core network' that responds more to viewing images of scenes rather than images of isolated objects (Zeidman et al., 2015). These regions were: Early Occipital cortex (OCC), left Lateral Occipital cortex (lLOC), left

Parahippocampal cortex (lPHC), left Retrosplenial cortex (lRSC), right Lateral Occipital cortex (rLOC), right Parahippocampal cortex (rPHC) and right Retrosplenial cortex (rRSC).

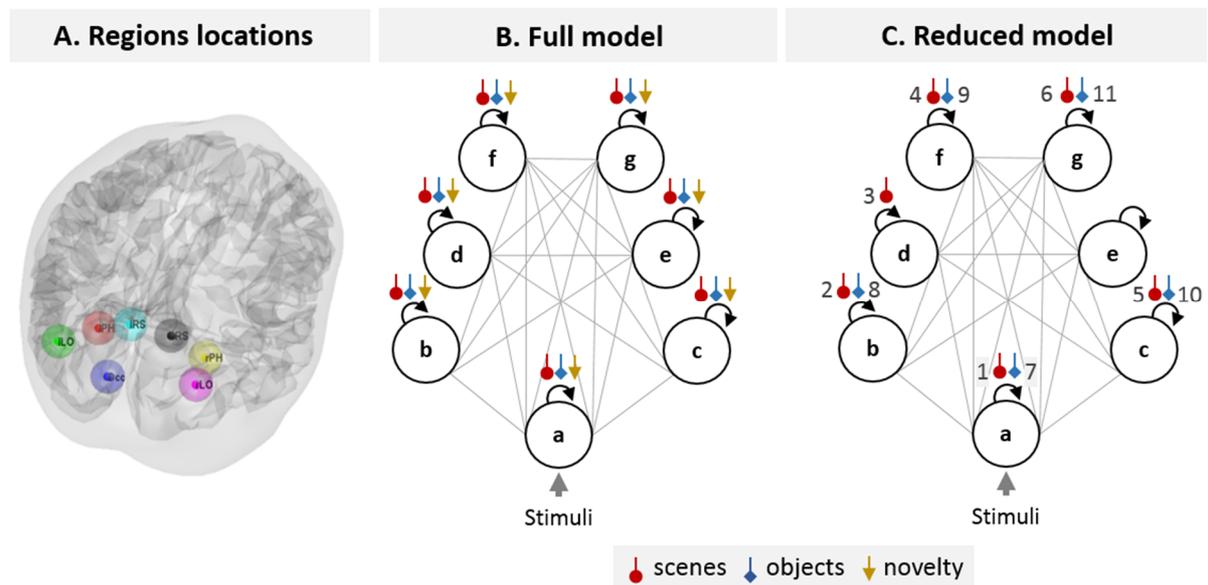

**Figure 3 DCM specification**. **A**. Locations of the 7 brain regions included in the DCM projected onto a canonical brain mesh. **B**. Structure of the DCM model estimated for each subject. The circles are brain regions, which were fully connected to one another (gray lines). The self-connection parameters (black arrows), which control each region's sensitivity to input from other regions, were modulated by each of the three experimental manipulations (coloured arrows). **C**. The optimal group-level (GLM) model after pruning away any parameters that did not contribute to the free energy. The numbered parameters correspond to the bar charts in Figure 4. Key: **a**=early visual cortex, **b**=left lateral occipital cortex, **c**=right lateral occipital cortex, **d**=left parahippocampal cortex, **e**=right parahippocampal cortex, **f**=left retrosplenial cortex, **g**=right retrosplenial cortex.

We extracted timeseries from each of these regions as follows. The group-level activation peak (collapsed across multiband factor to prevent bias) was identified from the contrast of scenes>objects (thresholded at $p < 0.05$ FWE-corrected) using a one-way ANOVA as implemented in SPM. Subsequently, a spherical region of interest (ROI) with 8mm FWHM was centred on the peaks at the individual level that were closest to the group-level peaks. This size of ROI sphere was arbitrary and provided a suitable trade-off between including a reasonable number of voxels and not crossing into neighbouring anatomical areas. Voxels within each sphere surviving at least $p < 0.001$ uncorrected at the single-subject level were summarised by their first principal eigenvariate, which formed the data feature used for subsequent DCM analysis.

The neuronal model for each subject's DCM was specified as a fully connected network (Figure 3B). Dynamics within the network were driven by all trials, modelled as boxcar functions, driving occipital cortex (the circle labelled **a** in Figure 3B). The experimental manipulations (scene stimuli, object stimuli and stimulus novelty) were modelled as modulating each region's self-inhibition (coloured arrows in Figure 3B). These parameters control the sensitivity of each region to inputs from the rest of the network, in each experimental condition. Neurobiologically, they serve as simple proxies for context-specific changes in the excitatory-inhibitory balance of pyramidal cells and inhibitory interneurons within each region (Bastos et al., 2012). These parameters, which form the B-matrix in the DCM neuronal model (Equation 3), are usually the most interesting from the experimenters' perspective – and we focussed on these parameters for our analyses.

## Results

We followed the analysis pipeline described above (see Summary of measures & analysis) to compare data acquired under four levels of multiband acceleration. The group-level results below and the associated figures were generated using the Matlab function **spm_dcm_bdc.m**.

### MB leakage / aliasing investigation

While not the focus of this paper, we conducted an analysis to ensure that our DCM results were not influenced by a potential image acquisition confound. As with any accelerated imaging technique, the multiband acquisition scheme is vulnerable to potential aliased signals being unfolded incorrectly. This is important since activation aliasing between DCM regions of interest could potentially lead to artificial correlations between regions (Todd et al., 2016). This analysis, detailed in the Supplementary Materials, confirmed that the aliased location of any given region of interest used in the DCM analysis did not overlap with any other region of interest.

### Identifying the optimal group-level model

We obtained estimates of each subject's neuronal responses by fitting a DCM to their data, separately for each multiband factor (the structure of this DCM is illustrated in Figure 3B). Then we estimated a single group-level Bayesian GLM of the neuronal parameters and pruned any parameters that did not contribute to the model evidence (by fixing them at their prior expectations with zero mean and zero variance). This gave the optimal group-level model, the parameters of which are illustrated in Figure 3C. The main conditions of interest were scene and object stimuli. Redundant modulatory effects of object stimuli were pruned from bilateral PHC, while the effect of scene stimuli was pruned from right PHC only. Redundant effects of stimulus novelty were pruned from all regions. This was not surprising, as the experimental design was not optimised for this contrast – and the regions of interest were not selected on the basis of tests for novelty effects.

### Modelling each dataset

Having identified a single group-level model architecture across all datasets (Figure 3C), we next updated each subject's DCMs to use this reduced architecture (by setting their priors to match the group level posteriors and obtaining updated estimates of the DCM parameters). We then estimated a group-level GLM for each dataset. The parameters of these four group-level GLMs are illustrated in Figure 4A-D. The numbered parameters, which correspond to those in Figure 3C, describe the change of sensitivity of each region to their inputs. More positive values signify more inhibition due to the task and more negative values signify dis-inhibition (excitation) due to the task. The results were largely consistent across multiband factors, with scene and object stimuli exciting most regions relative to baseline. Interestingly, modulation of early visual cortex by scenes and objects (parameters 1 and 7) were the largest effect sizes, so contributed the most to explaining the network-wide difference in scene and object stimuli.

Figure 4E shows the precision (inverse variance) of each parameter from Figure 4A-D. Each group of bars relates to a neuronal parameter, and each of the four bars relate to the four datasets (i.e. each of the four multiband factors). It is immediately apparent that all parameters (with the exception of parameter 7) achieved the highest precision with dataset MB4 (i.e. multiband acceleration factor 4). However, examining each parameter separately in this way is limited, because we cannot see the covariance between the parameters. The covariance is important in determining the confidence with which we can make inferences about parameters or models. Next, we apply our novel series of measures to these data, which provide a simple summary of the qualities of each dataset while taking into account the full parameter covariance.

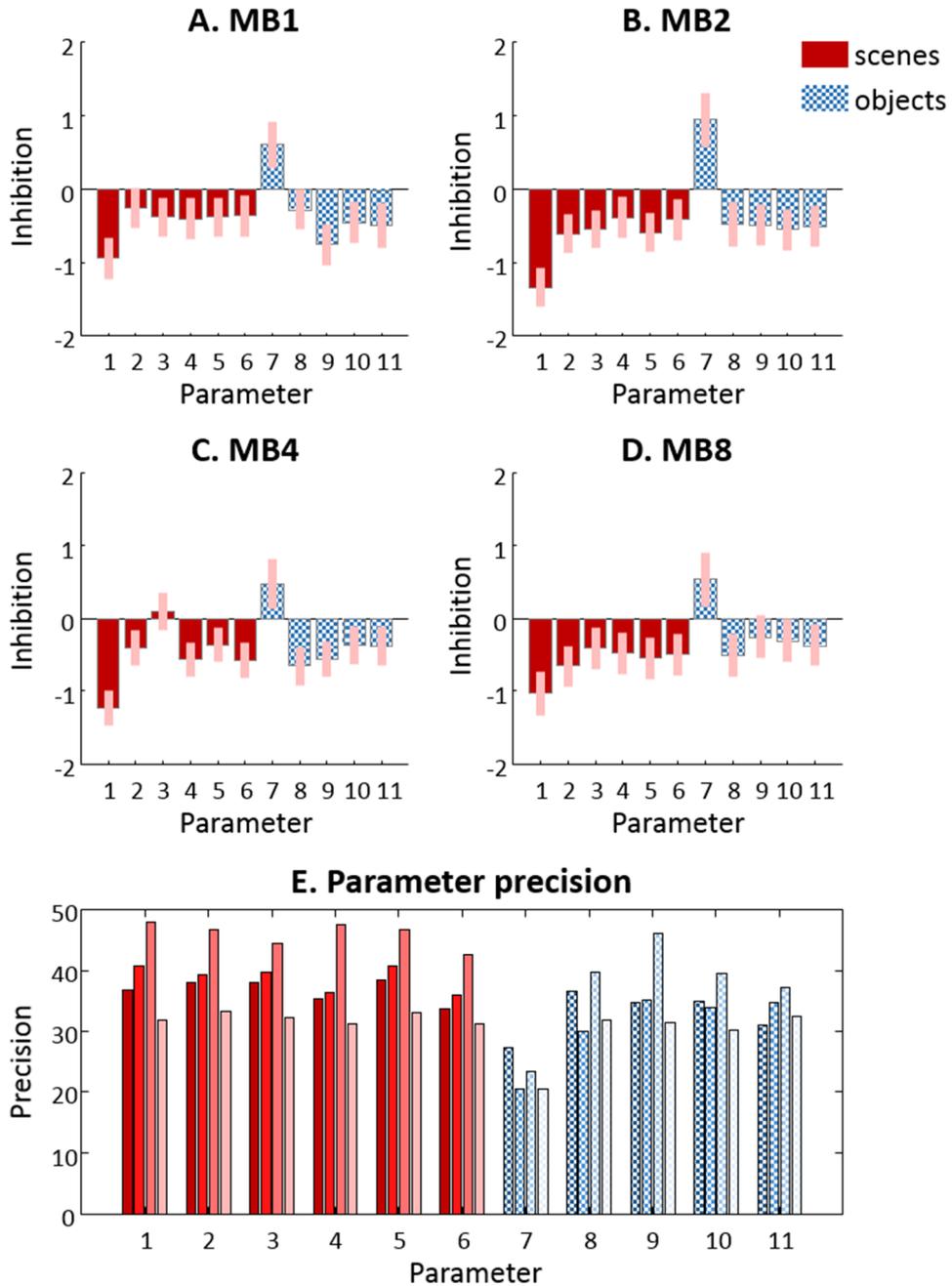

**Figure 4 Parameters of the group-level General Linear Model fitted to each dataset. A-D.** Posterior estimates of each parameter from each dataset. The bars correspond to the parameters labelled in Figure 3C, and for clarity these are divided into regional effects of scene stimuli (solid red) and of object stimuli (chequered blue). These parameters scale the prior self-connection of each region, and have no units. Positive values indicate greater inhibition due to the experimental condition and negative values indicate disinhibition (excitation). Pink error bars indicate 95% confidence intervals. MBx=multiband acceleration factor x. **E.** The precision of each parameter – i.e. the inverse of the variance which was used to form the pink error bars in Figure 4A-D. Each group of 4 bars denote the 4 datasets in the order MB=1, 2, 4 and 8 from left to right.

## Comparing datasets

In agreement with the analysis above, the dataset with multiband acceleration factor 4 (MB4) gave neuronal parameter estimates with the greatest precision or certainty (Figure 5A), followed by MB1 and MB2, and the least precision was for MB8. The difference between the best (MB4) and worst (MB8) performing datasets was 1.64 nats, equivalent to 84% probability of a difference (calculated by applying the softmax function to the plotted values). This may be classed as a trend or 'positive evidence' for MB4 over MB8 (Kass and Raftery, 1995), however the evidence was not strong enough to claim that MB4 was better than MB1 or MB2.

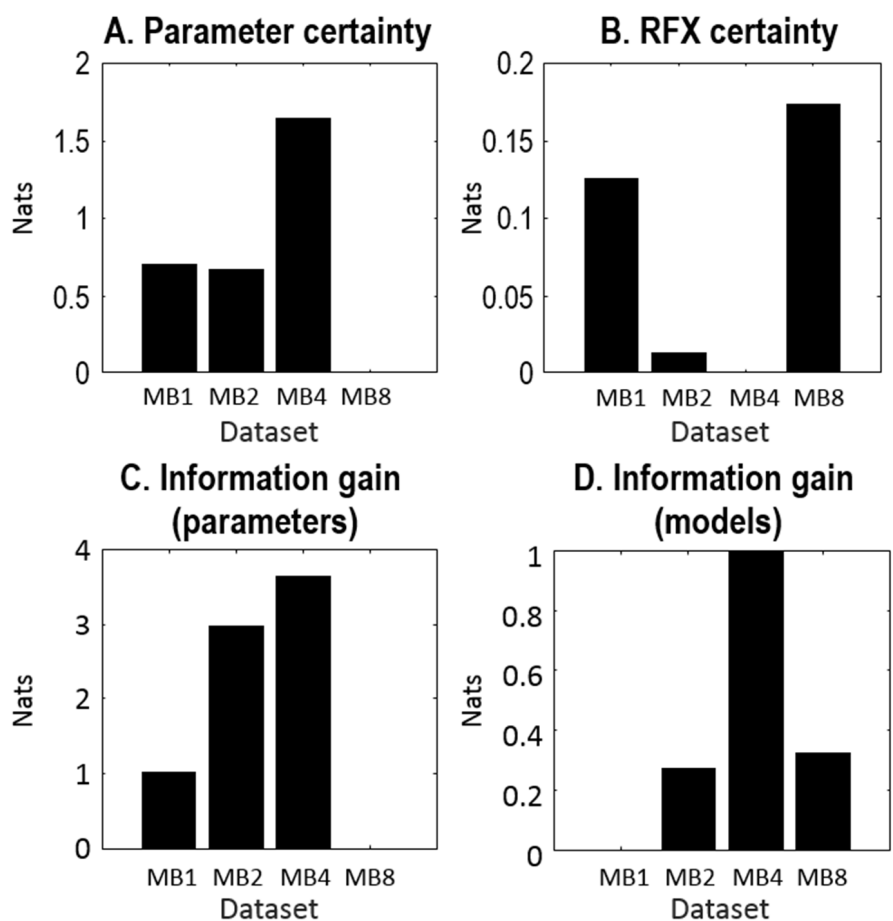

**Figure 5. Proposed measures for comparing datasets applied to empirical data.** For each measure, the y-axis is in units of nats relative to the worst model, which is set to zero. **A**. The relative negative entropy of the neuronal parameters of each dataset. **B**. The relative negative entropy of the estimated between-subjects variability. **C**. The information gain (KL-divergence) of the estimated neuronal parameters and the priors. **D**. The information gain (KL-divergence) between the estimated probability of each model and the prior belief that all models were equally likely. In each plot, the bars relate to 4 datasets which differed in their multiband (MB) acceleration factor: MB1, MB2, MB4, MB8.

Another parameter in the group-level Bayesian GLM (see Outcome measures, part b) quantified our belief about between-subject variability. The certainty about this parameter was very similar between datasets (Figure 5B). The difference between the best (MB8) and worst (MB1) datasets was 0.17 nats or 54% probability of a difference. This parameter was therefore not informative for selecting between datasets in this example.

The information gain over parameters (see Outcome measures, part c) is the extent to which the parameters were informed by the data. It reflects the number of independent parameters in the model (its complexity) which the data can accommodate. The best dataset was MB4 (Figure 5C), followed closely by MB2 and then MB1 and MB8. The difference between the best (MB4) and worst (MB8) datasets was 3.69 nats, or a 97.56% probability of a difference. There was also positive evidence that MB4 better than MB1 (2.65 nats = 93.41%), but MB4 could not be confidently distinguished from MB2 (0.69 nats = 66.66%). Thus, not only were the parameters most precise in dataset MB4 (Figure 5A), but they also gained the most information from the data, relative to the information available in the priors. This effect was particularly pronounced in comparison to MB8, and to a lesser extent, in comparison to MB1.

Finally we computed the information gain over models (see Outcome measures, part d), which quantifies the ability of the dataset to discriminate between similar models. The automated procedure described in the Methods section identified 8 similar Bayesian GLMs which differed in their priors (i.e. controlling which connectivity parameters were able to take on non-zero values). We found that MB4 afforded the best discrimination between these models (Figure 5D), however the effect size was small - the difference between MB4 and the worst dataset (MB1) was 0.99 nats, or a 73% probability of a difference, which is classed as 'weak evidence' (Kass and Raftery, 1995).

To summarise, three of the measures identified non-trivial differences between the datasets and each of these favoured the MB4 dataset. However, the magnitudes of these differences were generally not substantial. MB4 consistently performed better than MB8 – with positive evidence that MB4 provided more confident parameter estimates (Figure 5A) and greater information gain (Figure 5C). There was also positive evidence that MB4 offered greater information than MB1 (Figure 5C) and weak evidence that it supported greater information gain over models (Figure 5D). Given these results, if we were to conduct this same experiment with a larger sample, we would select multiband acceleration factor MB4 as our preferred acquisition protocol.

## Discussion

This paper introduced Bayesian data comparison, a systematic approach for identifying the optimal data features for inferring neuronal architectures. We have introduced a series of measures based on established Bayesian models of neuroimaging timeseries, in order to compare datasets for two types of analyses – inference about parameters and inference about models. We exemplified these measures using data from a published experiment, which investigated the performance of multiband fMRI in a cohort of ten healthy volunteers. This principled scheme, which can be applied by experimenters using a software function implemented in SPM (**spm_dcm_bdc.m**), can easily be applied to any experimental paradigm, any group of subjects (healthy or patient cohorts) and any acquisition scheme.

Comparing models based on their evidence is the most efficient procedure for testing hypotheses (Neyman, 1933) and is employed in both classical and Bayesian statistics. The model evidence is the probability of the data $y$ given the model $m$ i.e. $p(y|m)$. Model comparison involves taking the ratio of the evidences for competing models. However, this ratio (known as the Bayes factor) assumes that each model has been fitted to the same data $y$. This means that when deciding which data to use (e.g. arbitrating between different multiband acceleration factors) it is not possible to fit models to each dataset and compare them based on their evidence. To address this, the measures we introduced here can be used in place of the model evidence to decide which of several datasets provides the best estimates of model parameters and best distinguishes among competing models.

The first step in our proposed analysis scheme is to quantify neuronal responses for each data acquisition. This necessarily requires the use of a model to partition the variance into neuronal, haemodynamic and noise components. Any form of model and estimation scheme can be used, the only requirement being that it is probabilistic or Bayesian. In other words, it should furnish a probability density over the parameters. Here, we modelled each subject's neuronal activity using DCM for fMRI, in which the parameters form a multivariate normal distribution defined by the expected value of each parameter and their covariance (i.e., properly accounting for conditional dependencies among the parameters). Given that the main application of DCM is for investigating effective (causal) connectivity, the method offered in this paper is especially pertinent for asking which acquisition scheme will offer the most efficient estimates of connectivity parameters. Alternatively, the same analysis approach could be applied to the observation parameters rather than the neuronal parameters, to ask which dataset provides the best estimates of regional neurovascular coupling and the BOLD response. More broadly, any probabilistic model could have been used to obtain parameters relating to brain activity, one alternative being a Bayesian GLM at the single subject level (Penny et al., 2003).

Cognitive neuroscience research generally concerns effects that are conserved at the group level. However, the benefits of advanced acquisition schemes seen at the single subject level may not be preserved at the group level due to inter-subject variability (Kirilina et al., 2016). We were therefore motivated to develop a protocol to ask which acquisition scheme offers the best inferences at the group level, while appropriately modelling inter-subject variability. To facilitate this, the second step in our analysis procedure is to take the estimated neuronal parameters from every subject and summarise them using a group level model. Here we use a Bayesian GLM, estimated using a hierarchical (PEB) scheme. This provides the average (expected value) of the connectivity parameters across subjects, as well as the uncertainty (covariance) of these parameters. It additionally provides the free energy approximation of the log model evidence of the GLM, which quantifies the relative goodness of the GLM in terms of accuracy minus complexity. The key advantage of this Bayesian approach, unlike the summary statistic approach used with the classical GLM in neuroimaging, is that it takes the full distribution over the parameters (both the expected values and covariance) from the single subject level to the group level. This is important in assessing the quality of datasets, where the subject-level uncertainty over the parameters is key to assessing their utility for parameter-level inference. Together, by fitting DCMs at the single subject level and then a Bayesian GLM at the group level, one can appropriately quantify neuronal responses at the group level.

Having obtained parameters and log model evidences of each dataset's GLM, the final stage of our analysis procedure is to apply a series of measures to each dataset. These measures are derived from information theory and quantify the ability of the data to support two complementary types of inference. Firstly, inference about parameters involves testing hypotheses about the parameters of a model; e.g., assessing whether a particular neuronal response is positive or negative. A good dataset will support precise estimates of the parameters (where precision is the inverse variance) and will support the parameters being distinguished from one another (i.e. minimise conditional dependencies). We evaluated these features in each dataset by using the negative entropy of the parameters and the information gain. These provide a straightforward summary of the utility of each dataset for inference over parameters. A complementary form of inference involves embodying each hypothesis as a model and comparing these models based on their log evidence $\ln p(y|m)$. This forms the basis of most DCM studies, where models differ in terms of which connections are switched on and off, or which connections receive experimental inputs (specified by setting the priors of each model). We assessed each dataset in terms of its ability to distinguish similar, plausible and difficult-to-discriminate models from one another. This involved an automated procedure for defining a set of similar models, and the use of an information theoretic quantity – the information gain – to determine how well the models could be distinguished from one another in each dataset. This measure can be interpreted as the amount we have learnt about the models by performing the model comparison, relative to our prior belief that all models were equally likely.

To exemplify the approach, we compared four fMRI datasets that differed in their multiband acceleration factor. The higher the acceleration factor, the faster the image acquisition. This affords the potential to better separate physiological noise from task-related variance – or to increase functional sensitivity by providing more samples per unit time. However, this comes with various costs, including reduced SNR and increased temporal auto-correlations. These were acquired in the context of an established fMRI paradigm, which elicited known effects in pre-defined regions of interest. The conclusion of the original study (Todd et al., 2017), which examined the datasets under the same pre-processing procedures used here, was that a multiband acceleration factor between 4 (conservative) and 8 (aggressive) should be used. In the present analysis, the dataset acquired with multiband acceleration factor 4 (MB4) afforded the most precise estimates of neuronal parameters, and the largest information gain in terms of both parameters and models (Figure 5), although the differences between MB4 and MB2 were small. Our analysis of residual leakage artefact (Supplementary material) showed this result was not confounded by aliasing, a common issue with multiband acquisition. Given that these data were decimated so as to have equivalent numbers of samples, regardless of MB factor, our results suggest that the improved sampling of physiological effects provided by multiband acceleration counterbalanced the loss of SNR. Speculatively, MB4 may have been optimal in terms of benefiting from physiological filtering (a sufficiently high Nyquist frequency to resolve breathing effects), despite any reduction in SNR. MB2 may have performed slightly less well because it suffered from the penalty of reduced SNR, without sufficient benefit from the filtering of physiological effects. Any advantage of MB8 in terms of physiological filtering may have been outweighed by the greater reduction in SNR.

One should exercise caution in generalising this multiband result, which may not hold for different paradigms or image setups (e.g., RF coil types, field strength, resolution, etc.) or if using variable numbers of data points. Going forward, the effect of each of these manipulations could be framed as a hypothesis, and tested using the procedures described here. One interesting future direction would be to investigate the contribution of the two pre-processing steps: filtering and decimation. Our data were filtered to provide improved sampling of physiological noise and were subsequently decimated in order to maintain a fixed number of data points for all multiband factors under investigation. This ensured a fair comparison of the datasets with equivalent handling of temporal auto-correlations. The protocol described here could be used to evaluate different filtering and decimation options. One might anticipate that the increased effective number of degrees of freedom within the data would be tempered by increased temporal auto-correlations arising from more rapid sampling. A further specific consideration for the application of multiband fMRI to connectivity analyses is whether differences in slice timing across different acquisition speeds could influence estimates of effective connectivity in DCM. This hypothesis could be tested explicitly through Bayesian model comparison, by comparing DCMs with alternative settings of its inbuilt slice timing model (Kiebel et al., 2007). Here, for simplicity, we used the default setting, thereby aligning the onsets to the middle of each volume.

An important consideration – when introducing any novel modelling approach or procedure – is validation. The measures described here are simply descriptions or summary statistics of the Bayesian or probabilistic models to which they are applied. The measures themselves depend on two statistics from information theory - the negative entropy and the KL-divergence, which are ubiquitous in mathematics and statistical physics. These statistics do not require validation in and of themselves, just as the t-statistic does not need validation when used to compare fMRI datasets using the GLM. Rather, the models to which these measures are applied require validation before being used to decide between acquisition protocols or test neurobiological hypotheses. Here, we used two extant models from the neuroimaging community – DCM for fMRI and the Bayesian GLM implemented in the PEB framework. The face validity of DCM for fMRI has been tested using simulations (Friston et al., 2003; Chumbley et al., 2007), its construct validity has been tested using extant modelling approaches (Penny et al., 2004; Lee et al., 2006), and its predictive validity has been tested using intracranial recordings (David et al., 2008b; David et al., 2008a; Reyt et al., 2010). The PEB model and the associated Bayesian Model Reduction scheme is more recent and so far has been validated in terms of its face validity using simulated data (Friston et al., 2016) and its reproducibility with empirical data (Litvak et al., 2015). More generally, for any specific question of which dataset to select, the key validation question is whether the conclusions generalise to repeated acquisitions (test-retest reliability), to different brain regions and to different experimental protocols. These hypotheses can be tested using the measures introduced above. The exemplar multiband analysis included here, though congruent with a previous study (Todd et al., 2017), did not include validation of these sorts, and so the conclusions should not be generalised.

Practically, we envisage that a comparison of datasets using the methods described here could be performed on small pilot groups of subjects, the results of which would inform decisions about which imaging protocol to use in a subsequent full-scale study. Regions of interest would be selected for inclusion in the model which are known to show experimental effects for the selected task - based on an initial analysis (e.g. SPM analysis) and / or based on previous studies. The pilot analysis would ideally have the same design – e.g. model structure – as intended for the full-scale study. This is because the quality measures depend on the neuronal parameters of the specific model(s) which will be used by the experimenter to test hypotheses. Following this, we do not expect there exists a 'best' acquisition protocol in general for any imaging modality. Rather, the best dataset for a particular experiment will depend on the specific hypotheses (i.e. models) being tested, and the ideal dataset will maximise the precision of the parameters and maximise the difference in evidence between models. We anticipate that the protocol introduced here, implemented in the software accompanying this paper, will prove useful for experimenters when choosing their acquisition protocols.

# Appendix 1: Bayesian data comparison

In this appendix we illustrate why the entropy of the posterior over model parameters obtained from two different datasets, can be compared as if they were log Bayes factors. We first consider the comparison of different models of the same data. We then turn to the complementary application of Bayes theorem to the comparison of different data under the same model, in terms of their relative ability to reduce uncertainty about model parameters.

## Bayesian model comparison

To clarify model comparison in a conventional setting, consider two models $m_1$ and $m_2$ fitted to the *same* data $y$. The Bayes factor in favour of model 1 is the ratio of the model evidences under each model:

$$BF_1 = \frac{p(y|m_1)}{p(y|m_2)} \tag{13}$$

In place of the model evidence $p(y|m)$ we estimate an approximation (lower bound) – the negative variational free energy [this usually calls on variational Bayes, although other approximations can be used such as harmonic means from sampling approximations and other (Akaike or Bayesian) information criteria]:

$$F \approx \ln p(y|m) \tag{14}$$

As we are working with the log of the model evidence, it is more convenient to also work with the log of the Bayes factor:

$$\ln BF_1 = \ln p(y|m_1) - \ln p(y|m_2) \tag{15}$$
$$\approx F_1 - F_2$$

The log Bayes factor for two models is simply the difference in their free energies, in units of nats. Kass & Raftery (1995) assigned labels to describe the strength of evidence for one model over another; for example, 'strong evidence' requires a Bayes factor of 20 or a log Bayes factor of $\ln(20) \approx 3$. We can also transform the log Bayes factor to a posterior probability for one model relative to the other under uninformative priors over models: $p(m_1) = p(m_2) = p(m)$. By using Bayes rule and re-arranging, we find this probability is a sigmoid (i.e., logistic) function of the log Bayes factor:

$$p(m_1|y) = \frac{p(y|m_1)p(m)}{p(y)} = \frac{p(y|m_1)}{p(y|m_1) + p(y|m_2)} = \frac{1}{1 + \exp(-\ln \text{BF}_1)} \quad (16$$

The final equality shows that the logistic function of the log Bayes factor in favour model 1 gives the posterior probability for model 1, relative to model 2. When dealing with multiple models, the logistic function becomes a softmax function (i.e., normalised exponential function).

### Accuracy and complexity

The log evidence or free energy can be decomposed into accuracy and complexity terms:

$$F = \underbrace{\langle \ln p(y|m) \rangle_q}_{\text{accuracy}} - \underbrace{\text{KL}[q(\theta) \parallel p(\theta)]}_{\text{complexity}} \quad (17$$

Where $q(\theta) \approx p(\theta|y)$ is the posterior over parameters, $p(\theta)$ are the priors on the parameters and $\langle \cdot \rangle_q$ denotes the expectation under $q(\theta)$. This says that the accuracy is expected log likelihood of the data under a particular model. Conversely, the complexity scores how far the parameters had to move from their priors to explain the data, as measured by the KL-divergence between the posterior and prior distributions. This divergence is also known as a relative entropy or information gain. In other words, the complexity scores the reduction in uncertainty afforded by the data. We can now use this interpretation of the complexity to compare the ability of different data sets to reduce uncertainty about the model parameters.

### Comparing across datasets

Consider one model fitted to two datasets $y_1$ and $y_2$ with approximate posteriors over the parameters $q_1(\theta)$ and $q_2(\theta)$ for each data set. The corresponding complexity difference is:

$$\begin{aligned}
&(\text{KL}[q_1(\theta) \parallel p(\theta)] - \text{KL}[q_2(\theta) \parallel p(\theta)]) \\
&= \langle \ln q_1(\theta) \rangle_q - \langle \ln q_2(\theta) \rangle_q \\
&= H[q_2(\theta)] - H[q_1(\theta)] \\
&= \langle \ln \frac{q_1(\theta)}{q_2(\theta)} \rangle_q
\end{aligned} \quad (18$$

Where $H[q(\theta)]$ is the entropy of $q(\theta)$. Therefore, the reduction in conditional uncertainty (i.e., the difference in the entropies of the approximate posteriors) corresponds to the difference in information gain afforded by the two sets of data. Because this difference is measured in *nats*, it has the same interpretation as a difference in log evidence – or a log odds ratio (i.e., Bayes factor). The last equality above shows that the difference in entropies (or complexities) corresponds to the expected log odds ratio of posterior beliefs about the parameters. The entropies are defined for the multivariate normal distribution in Equation 9 of the main text.

## Acknowledgements

The Wellcome Centre for Human Neuroimaging (PZ, SMK, JKF, MFC) is supported by Wellcome grant number 539208.

## Conflicts of interest statement

The authors have no conflicts of interest to declare.

## References

Ashburner J, Friston KJ (2005) Unified segmentation. NeuroImage 26:839-851.

Bastos AM, Usrey WM, Adams RA, Mangun GR, Fries P, Friston KJ (2012) Canonical microcircuits for predictive coding. Neuron 76:695-711.

Buxton RB, Uludag K, Dubowitz DJ, Liu TT (2004) Modeling the hemodynamic response to brain activation. NeuroImage 23 Suppl 1:S220-233.

Cauley SF, Polimeni JR, Bhat H, Wald LL, Setsompop K (2014) Interslice leakage artifact reduction technique for simultaneous multislice acquisitions. Magnetic resonance in medicine : official journal of the Society of Magnetic Resonance in Medicine / Society of Magnetic Resonance in Medicine 72:93-102.

Chumbley JR, Friston KJ, Fearn T, Kiebel SJ (2007) A Metropolis-Hastings algorithm for dynamic causal models. NeuroImage 38:478-487.

David O, Wozniak A, Minotti L, Kahane P (2008a) Preictal short-term plasticity induced by intracerebral 1 Hz stimulation. NeuroImage 39:1633-1646.

David O, Guillemain I, Saillet S, Reyt S, Deransart C, Segebarth C, Depaulis A (2008b) Identifying neural drivers with functional MRI: an electrophysiological validation. PLoS biology 6:2683-2697.

Friston KJ (2011) Functional and effective connectivity: a review. Brain connectivity 1:13-36.

Friston KJ, Harrison L, Penny W (2003) Dynamic causal modelling. NeuroImage 19:1273-1302.

Friston KJ, Litvak V, Oswal A, Razi A, Stephan KE, van Wijk BCM, Ziegler G, Zeidman P (2016) Bayesian model reduction and empirical Bayes for group (DCM) studies. NeuroImage 128:413-431.

Kass RE, Raftery AE (1995) Bayes Factors. J Am Stat Assoc 90:773-795.

Kiebel SJ, Kloppel S, Weiskopf N, Friston KJ (2007) Dynamic causal modeling: a generative model of slice timing in fMRI. NeuroImage 34:1487-1496.

Kirilina E, Lutti A, Poser BA, Blankenburg F, Weiskopf N (2016) The quest for the best: The impact of different EPI sequences on the sensitivity of random effect fMRI group analyses. NeuroImage 126:49-59.

Kullback S, Leibler RA (1951) On information and sufficiency. Annals of Mathematical Statistics 22:79-86.

Larkman DJ, Hajnal JV, Herlihy AH, Coutts GA, Young IR, Ehnholm G (2001) Use of multicoil arrays for separation of signal from multiple slices simultaneously excited. Journal of magnetic resonance imaging : JMRI 13:313-317.

Lee L, Friston K, Horwitz B (2006) Large-scale neural models and dynamic causal modelling. NeuroImage 30:1243-1254.

Litvak V, Garrido M, Zeidman P, Friston K (2015) Empirical Bayes for Group (DCM) Studies: A Reproducibility Study. Frontiers in human neuroscience 9:670.

Neyman JP, Egon S (1933) On the Problem of the Most Efficient Tests of Statistical Hypotheses. Philosophical Transactions of the Royal Society A: Mathematical, Physical and Engineering Sciences 231:289-337.

Penny W, Kiebel S, Friston K (2003) Variational Bayesian inference for fMRI time series. NeuroImage 19:727-741.

Penny WD, Stephan KE, Mechelli A, Friston KJ (2004) Modelling functional integration: a comparison of structural equation and dynamic causal models. NeuroImage 23 Suppl 1:S264-274.

Reyt S, Picq C, Sinniger V, Clarencon D, Bonaz B, David O (2010) Dynamic Causal Modelling and physiological confounds: a functional MRI study of vagus nerve stimulation. NeuroImage 52:1456-1464.

Setsompop K, Gagoski BA, Polimeni JR, Witzel T, Wedeen VJ, Wald LL (2012) Blipped-controlled aliasing in parallel imaging for simultaneous multislice echo planar imaging with reduced g-factor penalty. Magnetic resonance in medicine : official journal of the Society of Magnetic Resonance in Medicine / Society of Magnetic Resonance in Medicine 67:1210-1224.

Spreng RN, Mar RA, Kim AS (2009) The common neural basis of autobiographical memory, prospection, navigation, theory of mind, and the default mode: a quantitative meta-analysis. Journal of cognitive neuroscience 21:489-510.

Stephan KE, Weiskopf N, Drysdale PM, Robinson PA, Friston KJ (2007) Comparing hemodynamic models with DCM. NeuroImage 38:387-401.

Stephan KE, Penny WD, Daunizeau J, Moran RJ, Friston KJ (2009) Bayesian model selection for group studies. NeuroImage 46:1004-1017.

Todd N, Moeller S, Auerbach EJ, Yacoub E, Flandin G, Weiskopf N (2016) Evaluation of 2D multiband EPI imaging for high-resolution, whole-brain, task-based fMRI studies at 3T: Sensitivity and slice leakage artifacts. NeuroImage 124:32-42.

Todd N, Josephs O, Zeidman P, Flandin G, Moeller S, Weiskopf N (2017) Functional Sensitivity of 2D Simultaneous Multi-Slice Echo-Planar Imaging: Effects of Acceleration on g-factor and Physiological Noise. Front Neurosci 11:158.

Xu J, Moeller S, Auerbach EJ, Strupp J, Smith SM, Feinberg DA, Yacoub E, Ugurbil K (2013) Evaluation of slice accelerations using multiband echo planar imaging at 3 T. NeuroImage 83:991-1001.

Zeidman P, Mullally SL, Maguire EA (2015) Constructing, Perceiving, and Maintaining Scenes: Hippocampal Activity and Connectivity. Cerebral cortex 25:3836-3855.

# Optimising data for modelling neuronal responses

## Supplementary Material

### Slice leakage / aliasing analysis

As with any accelerated imaging technique, the mulitband acquisition scheme is vulnerable to potential aliased signals not being unfolded correctly. This is important since activation aliasing between DCM regions of interest could potentially lead to artificial correlations between regions (Todd et al., 2016). Successful multiband (MB) reconstruction can be expected to produce the same percentage of false positive voxels in aliased and non-aliased voxels. To investigate this in our data, we created subject-specific masks corresponding to the aliased locations of all seven ROIs for a given MB factor. The ROIs were projected from normalized space back into native space using the inverse deformation field and the aliased locations for these voxels were then determined for MB 2, MB 4, and MB 8 based on their specific aliasing pattern (given MB factor and CAIPI-shift employed). These aliased locations were then projected back into normalized space using the forward deformation field. All seven ROIs therefore have [MB-1] aliased locations that are shifted diagonally along the slice and phase encode direction from the originally defined ROI. We first checked whether any of the DCM ROIs overlaped with an aliased region. Next, the following four measures were calculated: 1) $\alpha$: the total number of activated voxels in the aliased regions, 2) $\beta$: the total number of voxels in the aliased regions, 3) $\delta$: the total number of voxels activated elsewhere in the brain (i.e. outside the ROIs used in the DCM analysis or their aliased locations) and, 4) $\gamma$: total number of voxels elsewhere in the brain. A Wilcoxon signed-rank test was performed to test the hypothesis that the percentage of activated voxels in the alised regions ($\alpha/\beta$) was not significantly higher than the percentage of voxels activated elsewhere ($\delta/\gamma$).

For each MB factor considered, the aliased location of a given ROI used in the DCM analysis did not overlap with any other ROI being used in the analysis. In addition, the number of activated voxels in the aliased locations of the ROIs used in the DCM analysis was not significantly higher than in non-aliased locations outside the ROIs under consideration (Wilcoxon signed-rank paired test results were: $p=0.2783$ for MB=2, $p=1$ for MB=4 and $p=0.9678$ for MB8). In the case of MB=2, the peak number of aliased voxels was 81 corresponding to 6.8% in one subject (with an average of 2.4% across all ten subjects). The peak number of aliased voxels for MB=4 was 83 corresponding to 2.9 % in one subject (with an average of 0.9 % across all ten subjects). The peak number of aliased voxels for MB=8 was 77 corresponding to 2.5% in one subject (with overall average 0.6% across all ten subjects).

### References

Todd N, Moeller S, Auerbach EJ, Yacoub E, Flandin G, Weiskopf N (2016) Evaluation of 2D multiband EPI imaging for high-resolution, whole-brain, task-based fMRI studies at 3T: Sensitivity and slice leakage artifacts. NeuroImage 124:32-42.